# An experimental modal testing method for subcritical flow around a cylinder

Zhen Lyu[1], Jiaqing Kou[2] and Weiwei Zhang[1] *
1. School of Aeronautics, Northwestern Polytechnical University, Xi'an 710072, China
2. ETSIAE-UPM-School of Aeronautics, Universidad Politécnica de Madrid, Plaza Cardenal Cisneros 3, E-28040 Madrid, Spain

**Abstract:** Modal analysis of fluid flows is essential to understand flow physics and fluid-solid interaction mechanisms, and to implement flow control. Unlike unstable flow, the intrinsic attenuation of subcritical flow has led to failures to experimentally extract the subcritical flow modes clearly. To this end, this paper proposes a modal testing method for subcritical flows. Using Dynamic Mode Decomposition (DMD), dominate modes of flow around a cylinder at subcritical Reynolds numbers are extracted experimentally for the first time. The extracted structures and parameters of the modes are in good agreement with the numerical results in the literature. It is found that the first-order von Kármán mode can be observed at a Reynolds number as low as 19.3, which is very close to the lower boundary of subcritical vortex induced vibrations (VIV). This experimental observation elucidates the correlation between von Kármán mode and fluid-solid interaction instability in subcritical flows.

**Key words**： flow mode, DMD, subcritical flow, flow around a cylinder

## 1. Introduction

Modal testing in structural mechanics has been proven to be effective and useful to analyze the dynamic properties of structure [1]. In contrast, the extension of modal testing to fluid flows is more challenging due to the complicated temporal and spatial dynamic flow feature [2]. In recent years, several data-driven modal analysis techniques have been developed to overcome this challenge [3, 4]. The resulting flow modes are important to analyze physical mechanics and construct reduced-order flow models. Among these methods, Proper Orthogonal Decomposition (POD) [5] and Dynamic Mode Decomposition (DMD) [6] are the two most popular data-driven modal analysis methods.

POD is one of the first data-driven modal decomposition methods, which obtains relevant spatially orthogonal modes ranked by the mode energy [5]. However, it ignores the low-energy flow characteristics which may potentially have a large impact on flow dynamics. In contrast, DMD [6] extracts the flow mode from frequency content instead of energy. It obtains dominant flow modes with a single frequency and growth rate, which is consistent to structural modal analysis and characterize the stability behaviors. Hence, DMD shows a unique advantage in analyzing the periodic and quasi-periodic flows. Applications of POD and DMD to various complex flow problems were well documented by Taira [7].

The subcritical flow in the present study refers to intrinsically stable flows, e.g., flow past a circular cylinder at Re < 47 (above which periodic vortex shedding occurs). For such flows, the unsteady characteristics could not be observed without proper perturbation. At present, almost all research on modal analysis focuses on unstable flows [7], while modal analysis on subcritical flows remains limited. In fact, in fluid-structure interaction (FSI) and flow control, the flow mode of the subcritical flows also has a significant impact on the dynamic stability and response characteristics of the flow field. Bucci [8] investigated the onset of unsteadiness in a boundary-layer flow past a cylindrical roughness element at subcritical Reynolds numbers using experimental and numerical simulations. They found that the quasi-resonance of such a global mode to external forcing might be responsible for the onset of unsteadiness at subcritical Reynolds numbers. Gao et al. [9] examined the transonic buffet onset boundary numerically for an elastically mounted pitching NACA0012 airfoil. They found that the stability of the flow can be reduced due to the coupling between the subcritical flow mode and the structural mode in the pre-buffet state, which will result in a reduction of transonic buffet onset. Buffoni [10] investigated the wake of a forced oscillating cylinder experimentally and discovered that vortex shedding could be triggered at a Reynolds number as low as 25. Mittal et al. [11] numerically investigated the vortex induced vibration (VIV) of a cylinder at subcritical Reynolds numbers and observed that the lowest Re for the existence of subcritical VIV is 20. Using global linear stability analysis, they found that the subcritical VIV is caused by the instability of the FSI system. Subsequently, Kou [12] numerically studied the subcritical VIV of a cylinder with two degrees of freedom using a reduced-order model based stability analysis method. They found the lowest Reynolds number that incurs VIV is 18. Moreover, DMD was also performed by them to extract the purely flow modes of flow around a cylinder, with the DMD snapshots being acquired from direct numerical simulations. The stable von Kármán mode has been found to exist at subcritical Reynolds numbers and is almost invisible for Re < 18, indicating the lower boundary of VIV is Re = 18. Recently, Boersma et al. [14] conducted VIV experiments in a rotating channel, giving the first experimental evidence for the existence of subcritical VIV. In their experiments, VIV was observed to occur at a minimum Reynolds number of 19, which is very close to the numerical results of Mittal and Li. Almost simultaneously, Lyu et al. [15] carried out similar experiments in which they also observed subcritical VIV and came to similar conclusions.

Although subcritical flow modes and their significance for FSI and flow control have been discovered in the above numerical or experimental investigations, the exact structure of the subcritical flow mode has not been experimentally extracted. This is because subcritical flows are intrinsically stable and will decay fast to the steady state after a perturbation, making the modal testing very challenging. Therefore, this paper presents a modal testing method for subcritical flows. Through DMD, the dominant modes of flow around a cylinder at subcritical Reynolds numbers are extracted, and it was found that von Kármán shedding still dominants the flow dynamics at subcritical Reynolds numbers. Furthermore, the evolution of the flow mode along with Reynolds numbers was studied.

## 2. Principles and methods of flow modal testing

### 2.1 DMD method

As a data-driven modal analysis approach, DMD takes two snapshot sequences to extract the dynamic modes only from simulation data:

$$X = [x_1, x_2, x_3, ..., x_{N-1}] \text{ and } Y = [x_2, x_3, x_4, ..., x_N],$$

where $x_i$ is the $i$th snapshot and the time interval between two snapshots is fixed to $\Delta t$. Following Koopman theory, the nonlinear flow system can be approximated by an infinite-dimensional linear operator: $x_{i+1} = Ax_i$.

Then the relationship between snapshot X and snapshot Y can be obtained as

$$Y = [Ax_1, Ax_2, Ax, \cdots, Ax_{N-1}] = AX.$$

Since matrix A is usually high-dimensional, a similar matrix $\tilde{A}$ is constructed which has a lower dimension but approximates the main eigenvalues of the dynamic system. Singular value decomposition (SVD) is performed on X to determine this subspace

$$X = U\Sigma V^H, A = U\tilde{A}U^H.$$

where the singular value matrix $\Sigma$ contains $r$ non-zero singular values $\{\sigma_1, \cdots, \sigma_r\}$ in its diagonal. Since matrix $\tilde{A}$ has some of the eigenvalues of $A$, we can calculate the $j$th eigenvalue of matrix $\tilde{A}$, instead of calculating the eigenvalue of matrix A. The corresponding DMD mode is defined as $\Phi_j = Uw_j$, where $w_j$ is the eigenvector corresponding to the $j$th eigenvalue $\mu_j$. The growth rate $g_j$ and physical frequency $\omega_j$ of this mode are $g_j = \text{Re}\{\log(\mu_j)\}/\Delta t$ and $\omega_j = \text{Im}\{\log(\mu_j)\}/\Delta t$, respectively. More details about this algorithm are given in the work of Tu et al [16].

### 2.2 Modal testing methodology

In flow modal testing, snapshots that properly describe flow evolution are necessary. Hence, for subcritical flows, excitation that perturbs the flows to adequately obtain useful snapshots is the key in modal testing. It is found that the response amplifies greatly when the frequency of external excitation is close to the characteristic frequency of the flow. This phenomenon can be explained as "resonance". Therefore, the spectrum of the excitation signal should be broad enough to include the characteristic frequencies of the flow, while the amplitude of the excitation should be limited. Impulse excitation is chosen to perturb the flow due to its broad spectrum and simple operation. Flow chart of the proposed modal testing method is illustrated in figure 1, which consists of the following steps:

(a) Designing the excitation signal. To adequately perturb the flow, the impulse signal should be designed with proper spectrum covering the characteristic frequencies of subcritical flows. In addition, the amplitude of the excitation signal should be limited to avoid nonlinearity while maintaining reasonable signal-to-noise ratio (SNR) for sensor sampling.

(b) Exciting the flow field. The designed excitation signal was imposed on the test model. The type of excitation can be either the displacement or the microjet pulses.

(c) Collecting the snapshots of the flow field. After the impulse excitation, the snapshots describing the flow decaying were collected using time-resolved Particle Image Velocimetry (TR-PIV) or other synchronous multi-point / full-field measurement methods.

(d) Data pre-processing. As the flow decays fast, the number of snapshots collected in one test is small, and the SNR is low. Hence, pre-processing is performed using Variational Modal Decomposition (VMD) to detrend and denoise the measured data.

(e) Modal analysis of flow field. DMD was introduced to obtain the dynamic modes and the corresponding characteristic frequencies of the flow. In general, the DMD mode that dominants the decay frequency is typically the intrinsic flow mode. However, there are limitations in experimental environment, such as the insufficient number of snapshots due to the fast decaying of the flow, low SNR of the measurement data, or the influence of unphysical and spurious flow modes due to numerical noise. These issues were handled by the innovative screening method to determine physical modes in this paper, including two steps:

①. The characteristic frequencies of the flow were determined by the spectral analysis of the measurements at several points in the flow field.

②. The DMD modes whose frequencies are consistent with the characteristic frequencies of the flow field with a high mode ranking[19], are the physical modes.

Apart from impulse excitation, resonant excitation could also be used to obtain subcritical flow modes. In this method, the experiment model is forced to vibrate at different frequencies, and the response of the flow field is measured simultaneously. The frequency at which the response exhibits the greatest amplitude is the characteristic frequency of the flow field, and the relative flow mode is the dominant subcritical flow mode. However, this method requires a large number of tests and the damping of the flow modes are difficult to acquire, which is not considered in the present study.

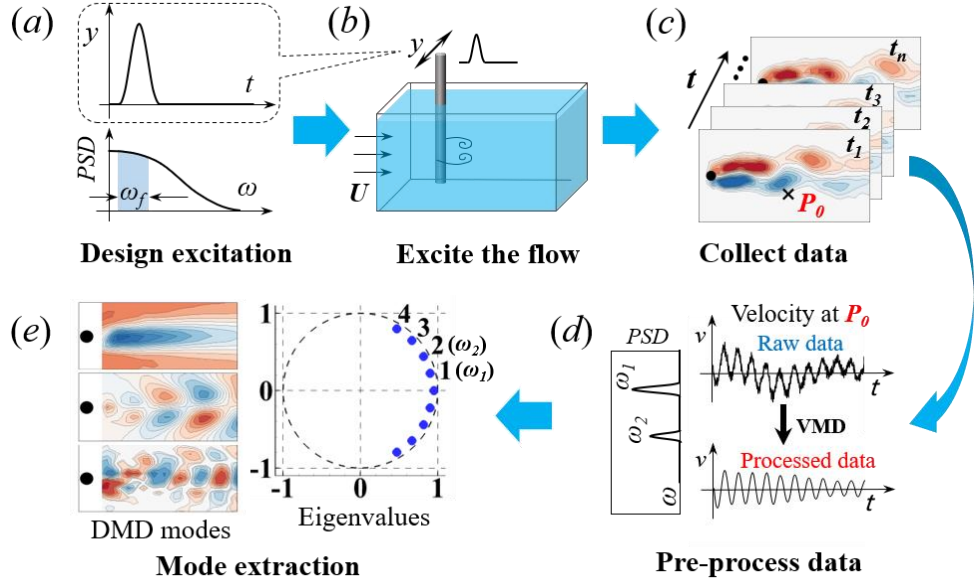

Figure 1. Flowchart of experimental modal testing for subcritical flow

## 3 Experiments of flow around a cylinder at subcritical Reynolds numbers

### *3.1 Rotating channel*

As the Reynolds number is very low in the present study, we must first create a steady flow at a sufficiently low velocity (a few centimeters per second). Given the difficulty of achieving such a low flow speed in a recirculating water channel, we designed a rotating channel by referring to the rotating tank in geophysical studies. A similar apparatus was also used in the experiment of Boersma et al [14]. As shown in Figure 2, the rotating channel mainly consists of two concentric cylinders with diameters of 500 mm and 380 mm, respectively. They were fastened to a transparent circular flat plate at one end. A stepper motor mounted in the center of the circular plate allows the channel to rotate at an angular speed range of 0 to 0.306 rad/s. The temperature of the water in the channel was strictly controlled at $18\pm0.1$ °C throughout the experiment to eliminate the effect of temperature on the viscosity coefficient.

We used planar PIV to quantify the speed distribution of the flowfield in the rotating channel. At a rotational speed of $\Omega$= 0.102 rad/s, Figure 2(c) depicts the instantaneous speed distribution of the horizontal cross-section of the channel. The flow speed is rather uniform, as can be seen in this figure. Figure 2(d) shows the turbulence intensity distribution in this state, revealing that turbulence in the center of the water tank is roughly 1%. To evaluate the effect of rotation, the Rossby number, defined as $Ro=U/2\Omega L=R/4d$, is introduced. Here, R is the distance between the model and the channel rotation axis, and d is the diameter of the model. From the Rossby number, the greater the effect of rotation on the flow, the smaller the Ro number is. In general, the effect of rotation on the flow is negligible when Ro > 1. In present study, Ro is in a range of 29.4-58.8, which is substantially larger than 1. Thus, the rotation effect on the flow can be neglected.

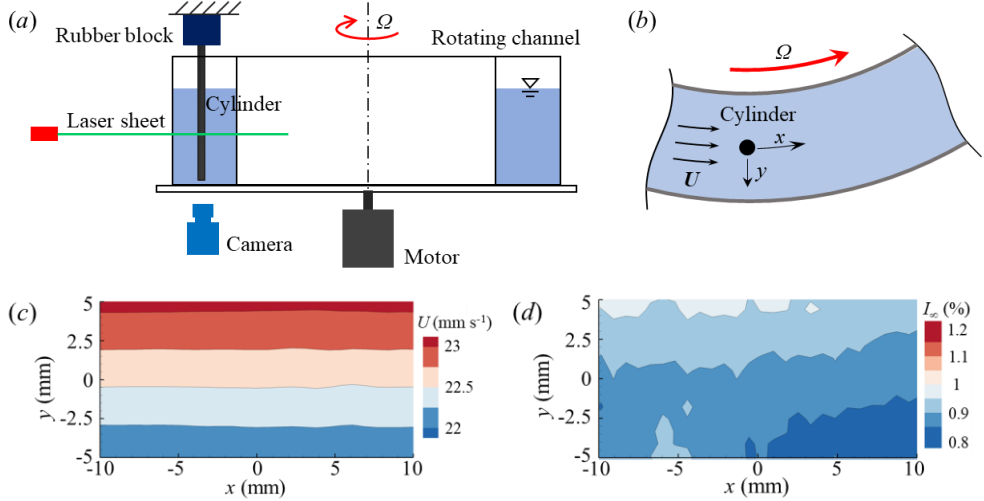

Figure 2. (a) Schematic of the experimental setup and (b) the definition of the coordinates. (c) Distribution of flow velocity and (d) turbulence intensity in the cross-section of the rotating channel at a speed of $\Omega = 0.102$ rad/s

### 3.2 Experimental set-up

The test model is an aluminum cylinder with a diameter of D = 2 mm and a length of L = 100 mm. The top of the model was held by a high-damping rubber block, as shown in figure 2(a). Planar time-resolved PIV was performed to measure the velocity fields of the cylinder. The rotating channel was seeded with polyamide resin particles of density, $\rho=1.03g/mm^3$ and mean diameter, d=20μm. A continuous laser was used to produce a laser sheet with a thickness of approximately 1mm. The laser plane was 45mm from the bottom of the channel. A high-speed camera (Pointgrey GS3-U3-23S6M-C) with 1920 ×1200 image resolution was used to capture the particle images. 9000 snapshots were captured at a sample rate of 100Hz for each single test case. The post-processing to acquire the velocity fields was performed with an open-source MATLAB package, PIVlab v2.3.1. The final interrogation window size used in the post-processing was 24 pixels × 24 pixels with an overlap ratio of 50%, resulting in a vector resolution of 0.38mm.

The experimental modal testing method introduced in Sec. 2.2 is applied here. We used a servo to tap the rubber block lightly at the bottom of the model in the direction perpendicular to the flow velocity. The temporal displacement response of the cylinder and its PSD are illustrated in figure 3. To improve the confidence level of results, the modal testing was repeated 20 times at each Reynolds number.

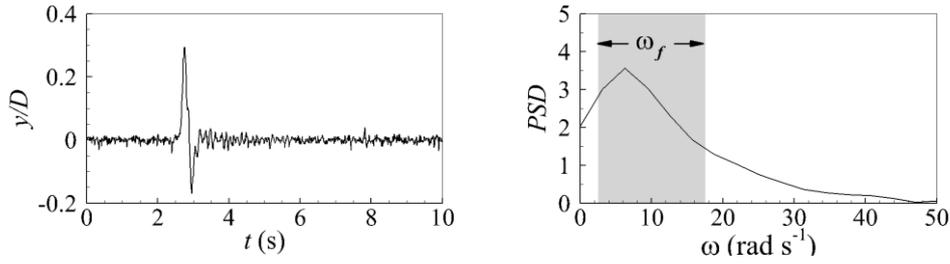

Figure 3. Model response after impulse excitation. (a) Temporal displacement response. (b) Spectrum of displacement response. $\omega_f$ denotes the range of characteristic frequencies of the flow. The spectrum is broad enough to include the characteristic frequencies of the flow.

## 4. Results and discussion

### 4.1 Flow around a cylinder at supercritical *Re*

To validate the feasibility and test the accuracy of the experimental set-up, we visualized the wake of a stationary cylinder at various Reynolds numbers, where the Strouhal numbers were also measured. Figure 4 shows the wake of the fixed cylinder at various Reynolds numbers. When Re < 47 (figures 4(a)), two shear layers are visible in the wake and do not interact with each other. No vortex shedding was observed behind the cylinder, indicating that the flow is stable. When Re < 47 (figures 4(b)), alternating shedding vortices can be observed behind the cylinder, indicating that the flow becomes unstable. These results indicate that the critical Reynolds number of the cylinder in the current experiments is approximately 47, which is consistent with the commonly accepted value of experiments and direct numerical simulations. Moreover, Figure 4(c) shows the Strouhal numbers for Re > 47. It can be seen that the current experimental results are in good agreement with those reported by Berger[17] and Friehe [18].

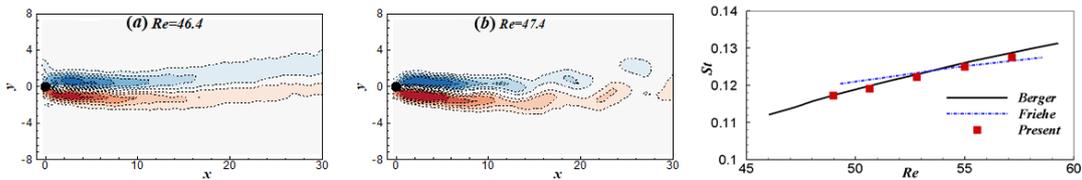

Figure 4. Wake patterns of a stationary cylinder at (a) Re=46.4 and (b) Re=47.4, indicating the critical Re is approximately 47. (c) Strouhal numbers of the stationary cylinder compared with Berger [17] and Friehe [18].

### 4.2 Modes of the flow around a cylinder at subcritical *Re*

The dominant modes of flow around a cylinder at Reynolds numbers ranging from 17 to 45 were captured. Figure 5(a) illustrates the transverse velocity decay responses for various subcritical Reynolds numbers following the same impulse excitation. As the Reynolds number decreases, the flow becomes more stable and the perturbation decays faster. Furthermore, Figure 5(b) and 5(c) depict the transverse velocity response and corresponding time-frequency spectrum at position (x, y) = (20D, 2D) for Re=40. It is observed that the dominant frequency of the flow field is approximately 1.15Hz and is related to the amplitude of the fluctuations in the flow field. Due to the nonlinearity of unsteady flows, the larger the amplitude of the fluctuation is, the higher the characteristic frequency is. This phenomenon can also be observed at supercritical Reynolds numbers where the frequency of vortex shedding at periodic limit cycle stage is higher than that at an unstable equilibrium state [19]. In addition, another frequency component of 2.3 Hz, which is the second harmonic of the dominant frequency, can be observed at t < 4.5 s. This implies that, in addition to the first order flow mode, a second order mode with a mode frequency twice the dominant frequency may exist. It should be emphasized that this mode exists only when the fluctuation of the flow field is larger, suggesting that the nonlinearity of the flow may be responsible for its formation. Such second order modes have been observed at supercritical Reynolds numbers. Based on numerical simulation, Taria [7] and Begheri [20] studied the modes of the flow around a cylinder at Re=100 by DMD and Koopman techniques, respectively, and discovered a second order flow mode with frequency twice the dominant frequency of the flow. Similarly, Hemati [21] experimentally studied the modes of flow around a cylinder at Re = 413 using the DMD method and found similar second order flow modes.

Modal analyses of the flow around a cylinder were performed at different subcritical and supercritical Reynolds numbers. The flow modes were distinguished from the DMD modes using the method described in Section 2.2, and the relevant mode frequencies and decay rates were obtained. Figure 6 depicts the Strouhal numbers and decay rates for flow around a cylinder at various subcritical Reynolds numbers. As shown in Figure 6(a), there is dispersion in the Strouhal number at each Reynolds number, due to the following two reasons: (1). As the Reynolds number decreases, the flow recovers to the steady-state more quickly. As a result, the number of valid snapshots recorded by TR-PIV becomes smaller, thus increasing the numerical error in the modal analysis. (2). To improve the SNR of the test data, the amplitude of the impulse excitation applied to the cylinder is slightly large, resulting in the nonlinearity of the flow. When the flow is nonlinear, its characteristic frequencies are related to the magnitude of the flow field fluctuations, as illustrated in Figure 5.

As shown in Figure 6(b), as the Reynolds number decreases, the damping ratio of the first order flow mode increases, indicating that the flow becomes more stable. For Re≥35, the relationship between the damping ratio and Reynolds number is nearly linear. Thus, interpolation of the damping ratios is performed, yielding a critical Reynolds number $Re_{cr}$ = 47.2, which is consistent with the experimental observations in section 4.1. Figure 6(c) shows the Strouhal number of the second order flow mode at various Reynolds numbers. In comparison to the first order flow mode, the second order flow mode is less intense and has a lower SNR, resulting in a larger Strouhal number dispersion. It should be emphasized that we were unable to observe the second order flow mode for Re < 30.8 regardless of the amplitude of the excitation applied to the cylinder. This implies that the occurrence of second order modes in flow around a cylinder requires a minimum Reynolds number of Re = 30.8.

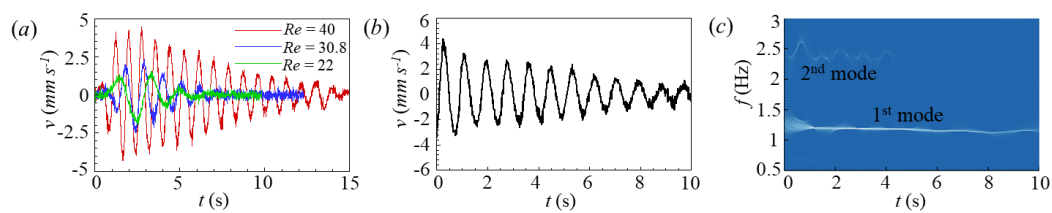

Figure 5. (a) Impulse excitation responses of the transverse velocity at *20D* downstream of the cylinder across various *Re*. (b) Transverse velocity response and (c) time-frequency spectrum at position *(x,y) = (20D, 2D)* for Re=40

Moreover, to validate the results of the modal analyses, Figure 6(d) depicts the Roshko number at various Reynolds numbers, where Ro is defined by the equation *Ro = StRe*. At subcritical Reynolds numbers, Ro increases linearly with Re and agrees well with the numerical results of Kou [20] and Chen [22].

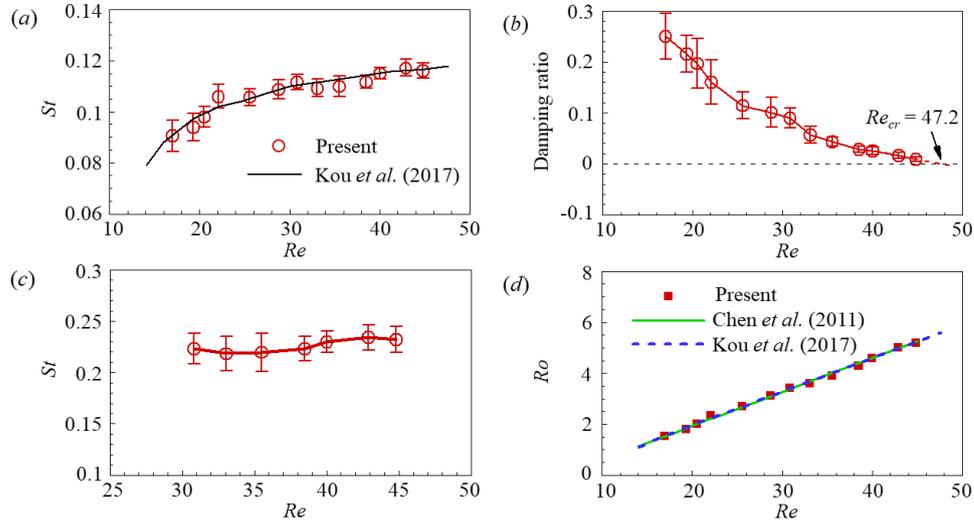

Figure 6. Parameters from the flow modal tests. (a) Strouhal number of 1st flow mode vs Re. (b) Damping ratio of 1st mode vs Re. (c) Strouhal number of 2nd flow mode vs *Re*. (d) *Roshko* numbers at subcritical *Re* compared with Kou [20] and Chen [22]. The error bars in above figures are standard deviations of the measured data from 20 identical modal tests.

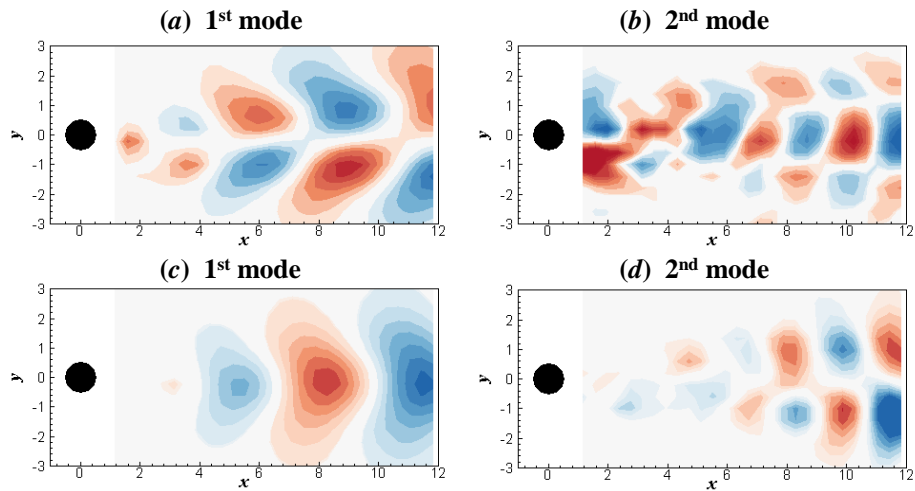

Figure 7. Dominant modes of flow around a cylinder at Re = 42.9. (*a*, b) are modes obtained from streamwise velocity snapshots, while (c, d) are modes obtained from the transverse velocity snapshots. The 1st flow mode is identical to the von Kármán mode in supercritical Re (see figure 9a).

Figure 7 illustrates the first and second order modes of flow around a cylinder at Re = 42.9. It can be seen that the structure of the first order flow mode at subcritical Reynolds numbers is identical to that of the von Kármán mode, indicating that the von Kármán mode can exist at subcritical Reynolds numbers. At supercritical Reynolds numbers, the von Kármán mode is unstable that incurs the well-known von Kármán vortex shedding. However, as illustrated in figures 6 and 7, the von Kármán mode at subcritical Reynolds numbers is stable and will decay to the steady-state after the external disturbance. This finding also verifies the numerical results of Kou[13] and Buffoni[10] Furthermore, the flow modes of the transverse velocity snapshot are substantially clearer and have a higher SNR than the flow modes of the streamwise velocity snapshot, and thus are employed in all of the following.

The normalized flow modes at Reynolds numbers 17, 19.3, 30.8, 35.5, 40, and 57.2 are shown in Figure 8. In particular, Figures 8(a-f) show the first-order modes of the flow field at different Reynolds numbers. As Re decreases, the spacing between adjacent vortices increases, indicating a decrease in St, which is consistent with the results in Figure 6(a). It should be noted that when Re is less than 19.3, the strength of the first order flow mode is significantly reduced, making this mode nearly invisible. This indicates the first order subcritical flow mode arises from Re=19.3, which is very close to the numerical result reported by Kou[13]. As shown in figure 8, the first order flow mode is an asymmetric mode that will contribute to the transverse aerodynamic forces. As pointed out by Kou et. al[13], the asymmetric flow mode could interact with the elastically supported structure, leading to the instability of the FSI system, which induces subcritical VIV. For Re<19.3, the first order flow mode nearly vanishes, making it extremely difficult for the FSI system to become unstable. This suggests that the lowest Reynolds number for the occurrence of subcritical VIV in experimental settings is likely to be 19.3, which is nearly identical to the experimental results of Boersma et al (i.e. Re=19). Additionally, Figures 8(g-j) illustrate the second order modes of the flow around a cylinder at various subcritical Re and their structure is comparable to that at a supercritical Re. Compared to the first order flow mode, the second order flow mode is less distinct and its SNR decreases as the Reynolds number decreases, eventually disappearing at Re < 30.8.

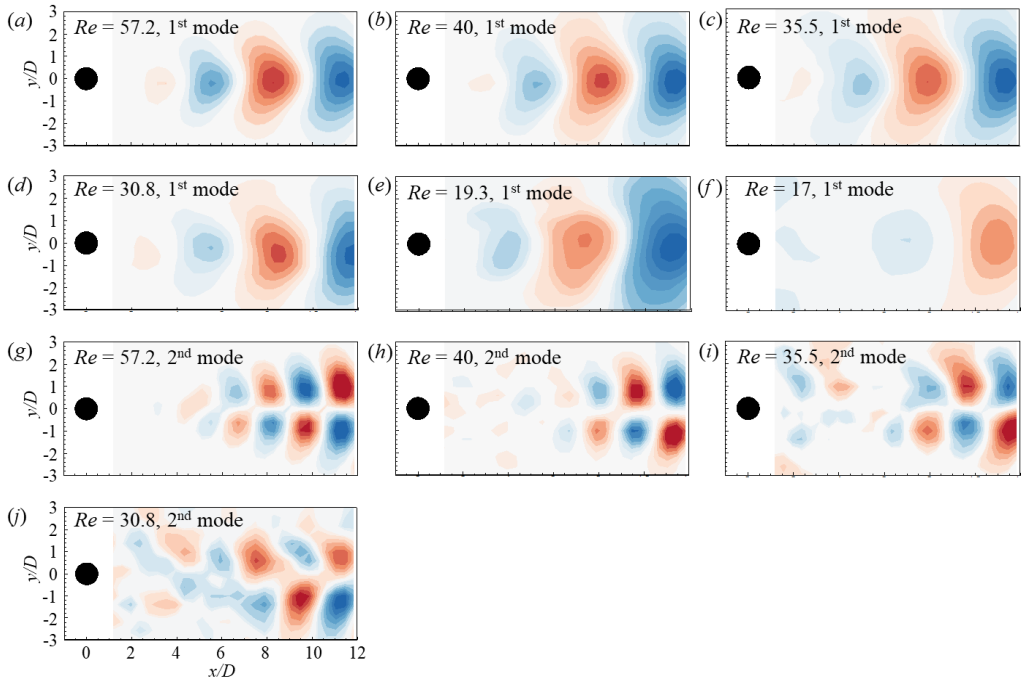

Figure 8. Dominant modes of flow around a cylinder at typical Reynolds numbers. (a-f) are 1st mode and (g-j) are 2nd mode. The von Kármán pattern in the 1st mode is nearly invisible for Re<19.3

## 5 Conclusion

An experimentally modal testing method for subcritical flow is proposed. Unlike the modal testing method for unstable flow, this method requires a well-designed excitation signal to perturb the flow field, as the subcritical flow is intrinsically attenuated. Using DMD, modal decomposition was performed to analyze flow data after an impulse excitation. The physical flow modes are filtered out from the DMD modes using the proposed modal screening approach, and the corresponding frequencies of the modes are determined simultaneously.

Subsequently, the proposed method has been validated through experimental studies on flow past a circular cylinder at subcritical Reynolds numbers. For the first time, the exact structures of the modes in subcritical flow were extracted experimentally. Results demonstrate that the structures, frequencies, and damping of the modes are consistent with numerical results of Kou. Moreover, it is found that the first order flow mode is von Karman mode and can be observed at a Reynolds number as low as 19.3. This Reynolds number is nearly identical to the lower boundary of the VIV (Re=19) observed experimentally by Boersma [14]. This finding allows to elucidate the association between von Kármán mode and FSI instability in subcritical flows.